\documentclass[11pt,final]{article}
\usepackage[color]{showkeys}

\usepackage{version}
\usepackage{amsmath}
\usepackage{graphicx}
\usepackage{amsthm}
\usepackage{amssymb}
\usepackage{alltt}
\usepackage{color}
\usepackage{epic}
\usepackage{eepic}
\usepackage{fancyheadings}
\usepackage{shadow}
\usepackage{array}
\usepackage{caption2}
\usepackage{enumerate}
\usepackage{ulem}
\usepackage{epsfig}
\usepackage{multicol}
\usepackage{ifthen}

\newcommand{\brk}[1]{\left(#1\right)}          
\newcommand{\fbrk}[1]{\!\left(#1\right)}       
\newcommand{\Brk}[1]{\left[#1\right]}          
\newcommand{\BRK}[1]{\left\{#1\right\}}        

\newcommand{\goto}{\rightarrow}

 {

\newtheorem*{prop*}{Proposition}
 }

\newcounter{sect}

\newcommand{\half}{\frac{1}{2}}

\renewcommand{\phi}{\varphi}

\newcommand{\MATLAB}{M\textsc{atlab }}

\renewcommand{\phi}{\varphi}
\renewcommand{\emph}[1]{{\it{#1}}}

\renewcommand{\Vec}[1]{{\bf{#1}}}

\title{The Tallest Column --- A Dynamical System Approach Using a Symmetry Solution}
\author{Yossi Farjoun\footnote{Department of Mathematics, UC
    Berkeley, Berkeley CA; e-mail: yfarjoun@math.berkeley.edu} \and John Neu\footnote{Department of Mathematics, UC
    Berkeley, Berkeley CA}}
\date{}

\begin{document}

\maketitle

\begin{abstract}

A classic problem, the design of the tallest column, is
solved again using a different method. 
By the use of a similarity solution the equations are transformed
and the difficult singularity at the endpoint is peeled away. 
The resulting autonomous system has a critical point and the solution
must be on its stable manifold. 
The solution is found by starting near the critical point in the
direction of the stable manifold, and solving backwards numerically.
This removes the need for an iterative integration method that was
previously used.
The method is shown to work for clamped or hinged boundary condition and
can also be used for other problems involving singularities at the endpoints. 
\end{abstract}


\section{Introduction}
\label{sec:intro}
We reexamine a classic eigenvalue maximization problem:
The design of the tallest column.
This problem has a lineage that originates from Euler's analysis of column buckling.
In \cite{KellerStrongest} Keller determines the tapering of a thin column with fixed volume which maximizes its buckling load.
Later, in \cite{KellerTallest}, Keller and Niordson solved the problem of designing the tallest free standing column.
The equations have a nasty singularity at the top of the column.
In \cite{KellerTallest} this singularity is neutralized by formulating an equivalent integral equation. 
The latter is solved by numerical iteration.

More recently, in \cite{CoxMcCarthy} and \cite{McCarthy}, Cox and
McCarthy showed that the operator involved can have a continuous
spectrum and they find the tapering of the tallest column  using other methods.
In this paper, no claims are made as to the validity of the equations involved or the derivation of them, see \cite{KirmserHu}, \cite{Stuart}, \cite{CoxMcCarthy} and \cite{McCarthy} for a discussion regarding the existence of the optimal design and the control requirements of the design. 
Nevertheless, Keller and Niordson's solution is highly plausible since the problem is infinitesimally close to a problem with a discrete spectrum.

The current paper concentrates on solving the difficult
equations (\ref{eq:ODEenergy}---\ref{eq:BC_at_tip}), originally derived by Keller and Niordson, by a new method.

First, in Section \ref{sec:deriv}, we derive the boundary value problem (BVP) which is an ordinary
differential equation (ODE) system for the deflection of the column
and the cross-sectional area, together with boundary conditions (BC). 
The ODE's contains the buckling load eigenvalue, to be determined as
part of the solution. 
This derivation follows the derivation in \cite{KellerTallest} closely
and is included for completeness.

In Section \ref{sec:solution} we solve the BVP.
First we show that the ODE's have a scaling
symmetry and an exact similarity solution which
has the desired asymptotic behavior at the tip of the column. 
This is where the singularity is.
The similarity solution suggests a transformation of variables that ``peels away the singularity'' at the tip.
This transformation yields ODE's for the new variables which form an autonomous system (AS).
In the new variables, the similarity solution is represented by a critical point of the AS.
Also, due to another symmetry, the eigenvalue disappears from the
equations and now appears only in the BC at the base of the column.
The transformed BC dictate that the solution must start on a certain
surface in the phase space of the AS and approach the critical
point.

Since the critical point of the AS has both stable and unstable manifolds,
solving the AS numerically from the base to the critical point
is difficult---the solution tends to deviate to a growing solution.
Instead, we examine the behavior of the AS near the critical point in
order to solve the system backwards, from the tip (the critical point) to the base.
By linearizing the AS around the critical point, we identify stable and unstable directions.
Starting near the critical point, on the stable manifold, the AS is
solved numerically, until the solution intersects the surface that the
BC defines. 
The point of intersection with the surface determines the
eigenvalue and the problem is solved. 
The numerical results agree with those of Keller and Niordson. 

In Section \ref{sec:other_bc} the problem is solved for other boundary
conditions. 
This is inexpensive once all the groundwork has been done.

This new solution method applies to other eigenvalue maximization problems whose equations have certain scale-invariant structure.
In a companion paper this method  determines the tapering of a javelin so the its first mode has the highest frequency of vibration (subject to length and volume constraint).

\section{Derivation of the Boundary Value Problem}
\label{sec:deriv}
\subsection{Setup} 
\begin{figure}[ht!]
\begin{center}
\scalebox{0.50}{\includegraphics{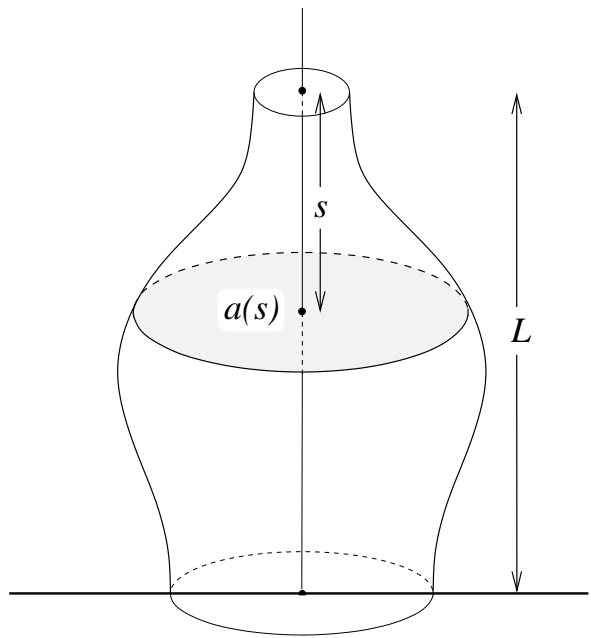}}
\caption{The shape of the column is governed by the cross-sectional
  area function $a(s)$ measuring the area of a cross-section at a
  point located at arclength $s$ measured along the center axis of
  the column from its tip. 
All the cross-sections are convex, are geometrically similar, and are equally oriented.}
\label{fig:alloweddesigns}
\end{center}
\end{figure}

Consider columns, all of the same volume, clamped at the base and free at the top.
We want to find the shape of the tallest column that will not buckle under its own
weight. 
As in the Keller and Niordson papers, we  solve the problem for a specific class of permissible designs.
We assume that the column is thin, i.e. the characteristic width is much less than the height of the column.
In addition, we only allow columns with geometrically similar, equally oriented and convex cross-sections. 
See figure (\ref{fig:alloweddesigns}).

\begin{figure}[ht]
\begin{center}
\scalebox{0.75}{\includegraphics{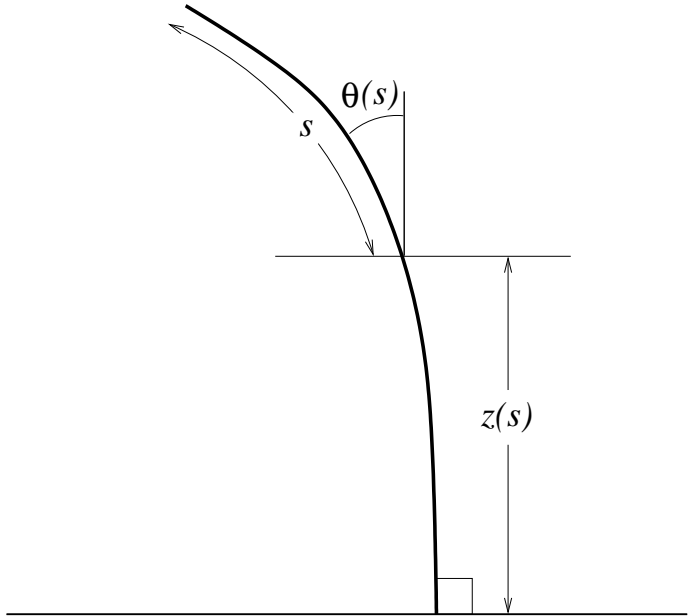}}
\caption{The shape of the bent column is specified by the angle that
  the center axis of the column makes with the vertical at each point. 
The points on the column are parameterized by arclength $s$, measured
  from the tip along the columns center axis. 
$z(s)$ is the vertical height of the point $s$.}
\label{fig:variables}
\end{center}
\end{figure}

We parameterize the column by arclength $s$, measured from the tip of
the column, along its center axis.
The design information is contained  in a single function $a(s)$,
 the cross-sectional area at $s$.
Lastly, we concern ourselves with the bending of the column in a specified plane only.
This allows us to specify the configuration of the column using a
single function $\theta(s)$---the angle that the center axis of the
column makes with the vertical, measured at point $s$.
See Figure (\ref{fig:variables}).

To find the BVP satisfied by $\theta(s)$ for a given cross-section
 $a(s)$, we write the total energy of the system as a
functional of $\theta(s)$, and compute its variational equations.

\subsection{Energy Minimization}
The total energy of the column, elastic and gravitational, is a
functional of $\theta$:
\begin{equation}
e[\theta] =\int_0^L\half b\fbrk{s}\theta_s^2(s)\,ds + g\rho\int_0^La\fbrk{s}z\fbrk{s}\,ds.
\label{eq:energy}
\end{equation}
Here, $L$ is the length of the column, $g$ is the gravitational acceleration, $\rho$ is the density of the material, and $z\fbrk{s}$ is the vertical elevation at $s$.
The bending modulus, $b(s)$ is proportional to $a^2(s)$,
\begin{equation}
b\fbrk{s} = cEa^2\fbrk{s}.
\label{eq:b}
\end{equation}
Here, $c$ is a dimensionless constant determined by the cross-sectional shape and orientation, and $E$ is Young's modulus.
For a discussion of the choice of cross-sectional designs and derivation of the bending modulus, see \cite{KellerStrongest}. 
Since $z(s)$ is the vertical elevation of a particle at $s$,
\begin{equation}
z\fbrk{s} = \int_s^1\cos\theta\fbrk{t} \,dt.
\label{eq:z}
\end{equation}
Substituting (\ref{eq:b},\ref{eq:z}) into (\ref{eq:energy}) we get
\begin{equation}
e[\theta] =\int_0^L\BRK{\half cE\,a^2\fbrk{s}\theta_s^2(s) + g\rho\, a\fbrk{s}\int_s^1\cos\theta\fbrk{t}\,dt}\,ds.
\label{eq:energy2}
\end{equation}
We adopt dimensionless variables in which $s$, $a$ and $e$ are measured in the units given by the scaling table:
\begin{center}
\begin{tabular}{lccc}
Variable &$s$&$a$&$e$\\
\hline
Unit&$L$&$\frac{V}{L}$&$\frac{cEV^2}{L^3}$
\end{tabular}
\end{center}
Here, $V$ is the total volume of the column.
$V$ is related to $a(s)$ by
\begin{equation}
  V\equiv\int_0^L a(s)\,ds.\label{eq:volume}
\end{equation}
The dimensionless versions of (\ref{eq:energy2}) and (\ref{eq:volume}) are
\begin{align}
e[\theta] &=\int_0^1\BRK{\half {a}^{2}\fbrk{s}{\theta}_{s}^2(s) + \lambda\, a\fbrk{s}\int_s^1\cos\theta\fbrk{t}\,dt}\,ds,
\label{eq:energy3}\\
  V[a]&\equiv\int_0^1 a\fbrk{s}\,ds = 1
\label{eq:dimlessvolume}
\end{align}
Where $\lambda =\frac{\rho g L^3}{VcE}$ is the dimensionless load per unit volume.

In order to determine the variational BVP for $\theta$ associated with
the energy (\ref{eq:energy3}), it is convenient to change the
order of integration in the second term of equation
(\ref{eq:energy3}). Additional rearrangements give
\begin{align}
e\Brk{\theta} &=\int_0^1\BRK{\half a^2\fbrk{s}\theta_s^2(s) +
\lambda \cos \theta\fbrk{s}\int_0^s a\fbrk{t}\,dt} \,ds.
\label{eq:lagrangian2}
\end{align}

The variational BVP of \eqref{eq:lagrangian2} consists of the ODE 
\begin{equation}
 \brk{a^2(s)\,\theta_s(s)}_s + \lambda \sin \theta\fbrk{s}\int_0^sa\fbrk{t}\,dt =0
\label{eq:BVP}
\end{equation}
in $0<s<1$ and BC
\begin{align}
\theta(1) &= 0, & a^2\,\theta_s&=0\quad \text{at } s=0.
\label{eq:BC}
\end{align}
Physically, the BC mean that there is a zero angle at the
base of the column (clamped) and zero torque at the top (free).
Equation \eqref{eq:BVP} is in fact a local torque balance. An explanation
of this is given in Appendix \ref{app:torque}.

\subsection{Buckling Load}
Recall that $\lambda$ is the dimensionless load per unit volume. 
For a column tapering given by $a(s)$, there is a critical value $\lambda_c$ such that for $\lambda < \lambda_c$ the only solution of the BVP is $\theta(s)\equiv 0$.
For $\lambda>\lambda_c$ there are non-trivial solutions as well. 
This is the buckling phenomenon.
For such $\lambda$, a small perturbation of the column from the
undeflected state can have a smaller energy than the undeflected one, therefore
 the zero solution is no longer an energy minimum. 
The energy minimum is one of the non-zero solutions.
For $\lambda$ \emph{slightly} greater than $\lambda_c$ the energy minimizing solution  $\theta(s)$ will be small.
In this case we can study the linearization of
(\ref{eq:BVP}) about $\theta\equiv 0$.
The linearization of the BVP (\ref{eq:BVP},\ref{eq:BC}) is a
Sturm-Liouville eigenvalue problem:
\begin{gather}
 \brk{a^2\,\theta_s}_s + \lambda_c \theta\int_0^sa\fbrk{t}\,dt =0, \qquad \text{in } 0\le s\le 1
\label{eq:linBVP}\\
 \theta\fbrk{1}=0 \qquad a^2\,\theta_s=0 \text{ at } s=0.
\label{eq:linBC}
\end{gather}

The following analysis assumes, as in \cite{KellerTallest}, that the Sturm-Liouville problem determines a discrete sequence of
eigenvalues $\lambda$, and corresponding eigenfunctions which are
determined up to a multiplicative constant.

\subsection{Eigenvalue Maximization} 
What does the tallest column look like? 
That is, what is the distribution $a(s)$ of cross-sectional area which maximizes the smallest eigenvalue $\lambda_c$ of the linearized BVP (\ref{eq:linBVP}) subject to the volume constraint (\ref{eq:dimlessvolume})?

Since we have no other $\lambda$ but $\lambda_c$ we will drop the subscript and use $\lambda$ to mean the lowest eigenvalue of the linearized BVP.
$\lambda=\lambda[a]$ is a functional of $a(s)$. 
To maximize $\lambda[a]$ subject to the volume
constraint, we find the variational derivative
$\frac{\delta\lambda}{\delta a}$ and solve for the function $a(s)$ so that 
\begin{equation}
  \frac{\delta\lambda}{\delta a} = \mu \frac{\delta V}{\delta a}=\mu.
\label{eq:maximization}
\end{equation}
Here, $\mu$ is a Lagrange multiplier associated with the volume constraint (\ref{eq:dimlessvolume}).

To find $\frac{\delta\lambda}{\delta a}$, we introduce a small
variation in the cross-sectional area,
$\delta a(s)$. 
As a result, $\lambda$ and $\theta(s)$ will have corresponding small variations $\delta \lambda$ and $\delta\theta(s)$ respectively.
The linearization of equation (\ref{eq:linBVP}) in $\delta a,
\delta\theta$ and $\delta\lambda$ about a solution $a, \theta,
\lambda$, results in equation (\ref{eq:linInhom}). 
To reduce clutter, the $(s)$ notation is dropped whenever the
variable in the parenthesis is $s$, so $a, \delta a, \theta, \delta\theta$ mean  $a(s), \delta a(s),
\theta(s), \delta\theta(s)$ respectively.
Whenever other variables are used they are written explicitly.
\begin{equation}
\brk{2a\,\delta a\,\theta_s + a^2\,\delta\theta_s}_s+
\delta\!\lambda\,\theta\int_0^sa(t)\,dt+
\lambda\,\delta\theta\int_0^sa(t)\,dt+
\lambda\,\theta\,\int_0^s\delta a(t)\,dt=0.
\label{eq:linInhom}
\end{equation}
The BC for $\delta a, \delta\theta$ are
\begin{equation}
\delta\theta(1)=0 \qquad \brk{2a \,\delta a\, \theta_s
  +a^2\,\delta\theta_s}=0 \text{ at } s=0.\notag
\end{equation}
This is a linear inhomogeneous BVP for $\delta\theta$.
Its solvability condition determines the functional derivative $\frac{\delta\lambda}{\delta a}$.
To find the solvability condition, we multiply both sides of
(\ref{eq:linInhom}) by $\theta$ and integrate over $0\le s\le1$. 
Integration by parts, use of BC and rearrangement, leads to  
\begin{equation}
\int_0^1\BRK{ \brk{a^2\theta_s}_s+\lambda\theta\int_0^sa(t)\,dt}\delta\theta\,ds = \int_0^1\BRK{2a\,\delta a\,\theta_s^2 -
\delta\!\lambda\,\theta^2\int_0^sa(t)\,dt-
\lambda\,\delta a\int_s^1\theta^2(t)\,dt }\,ds.\notag
\end{equation}
The LHS is zero due to (\ref{eq:linBVP}), and
we get an equation relating $\delta a$ and $\delta\lambda$:
\begin{equation}
\int_0^1\BRK{2a\,\theta_s^2 -\lambda\int_s^1\theta^2(t)\,dt }\delta a \,ds = 
\delta\lambda\int_0^1\theta^2(r)\int_0^ra(t)\,dt\,dr.
\label{eq:funcder1}
\end{equation}
Here the integration variable in the last term on the right was
changed from $s$ to $r$ for added clarity.
In functional derivative form, equation (\ref{eq:funcder1}) becomes
\begin{equation}
\frac{\delta\lambda}{\delta a} = \frac{2a\,\theta_s^2 -\lambda\int_0^s\theta^2(t)\,dt }{\int_0^1\theta^2(r)\int_r^1a(t)\,dt\,dr}.\notag
\end{equation}
Substituting this into (\ref{eq:maximization}) results in an integro-differential equation for the maximizing cross-sectional area function $a$ and the critical load $\lambda$:
\begin{equation} 
2a\,\theta_s^2 -\lambda\int_s^1\theta^2(t)\,dt =\mu\int_0^1\theta^2(r)\int_0^ra(t)\,dt\,dr. \label{eq:evMax}
\end{equation}
Equation (\ref{eq:evMax})  must be solved together with (\ref{eq:linBVP}, \ref{eq:linBC}) to produce $a(s), \theta(s),$ and $\lambda$.
A short calculation (see Appendix \ref{calc:mu}) shows that $\mu=\lambda$.

The integro-differential system (\ref{eq:linBVP}, \ref{eq:linBC}, \ref{eq:evMax}) is still not in a form that we can easily solve. 
It is convenient to convert it into a system of ODE with BC.
Differentiating (\ref{eq:evMax}) with respect to $s$ gives the ODE
\begin{equation*} 
2\brk{a\,\theta_s^2}_s +\lambda\theta^2  = 0.
\end{equation*}
In order to transform equation (\ref{eq:linBVP}) into a ODE and get
rid of the integral, the variable $b(s)$ is introduced:
\begin{align}
  b &\equiv \int_0^s a(t)\,dt.\label{def_b2}
\end{align}
Differentiating \eqref{def_b2} gives a BVP for $b(s)$:
\begin{align}
  b_s = a, \qquad b(0)=0. \label{eq:b_BVP}
\end{align}

Physically, $b$ is the amount of volume above the point $s$.

\section{Solution of the BVP}
\label{sec:solution}
The system that we want to solve is
\begin{align}
 (a^2\theta_s)_s +\lambda b \theta &=0,    \label{eq:ODEenergy}\\
  2(a\theta_s^2)_s + \lambda \theta^2 &=0,  \label{eq:ODEmax}\\
  b_s - a &=0,                               \label{eq:ODEint}
\end{align}
with boundary conditions
\begin{align}
 \theta(1)&=0,& b(1)&=1,\label{eq:BC_at_base}\\
 a^2\,\theta_s&=0 \text{ at } s=0, & b(0) &=0 \label{eq:BC_at_tip}.
\end{align}
Heuristically, the optimal column is expected to taper to a point as $s$ approaches 0, so $a(s)$ is expected to go to zero as $s\rightarrow 0$.
In addition, $b(0)=0$ by the BC \eqref{eq:b_BVP}.
Since $a$ and $b$ both multiply equation (\ref{eq:ODEenergy}), one
expects severe difficulties in the numerical solution as $s$
approaches 0, and this is indeed the case.

This motivates an analysis of the asymptotic structure of the solution to (\ref{eq:ODEenergy}--\ref{eq:BC_at_tip}) as $s$ approaches 0.
We need to peel away the singularity before attempting any numerical solution.
In the following sections we use a similarity solution to remove this
singularity, transforming the equations so that they can be solved
numerically.
 
\subsection{Similarity Solution}
Equations (\ref{eq:ODEenergy}--\ref{eq:ODEint}) have a similarity
solution.
(For information on similarity solutions see, for example,
\cite{Barenblatt} or \cite{Bluman}.)
To find it we study the scaling relations between the variables:
Specifically, let $A, B, S$ and $T$ be ``units'' of $a, b, s$ and
$\theta$ respectively. 
A balance of units in  ODE's (\ref{eq:ODEenergy}--\ref{eq:BC_at_base}) gives the relationships
\begin{align*}
A^2TS^{-2}&=TB,\\
AT^2S^{-3} &= T^2,\\
B S^{-1} &= A.
\intertext{$T$ cancels out, and we are left with three equations for $A$, $B$ and $S$.
In fact, one equation is redundant, and we get a simple relation between $A, B$ and $S$:}
B &= S^4,\\
A &= S^3.
\end{align*}
Hence, it is expected that the ODE's are invariant under a scaling
transformation, so that if $\theta(s), a(s), b(s)$ are solutions, then so
are $\theta(s/S), S^3a(s/S), S^4b(s/S)$ for any $S$.
This suggests a similarity solution of the ODE with
\begin{align}
  \tilde a(s) &= a_0 s^3,\label{eq:ansatz_a}\\
  \tilde b(s) &= b_0 s^4.\label{eq:ansatz_b}
\intertext{
We also look for a similar behavior for $\theta$:}
\tilde\theta(s) &= \theta_0 s^p.\notag
\intertext{Since the equations are linear in $\theta$, we can choose $\theta_0=1$, therefore}
  \tilde\theta({s}) &= s^p.\label{eq:ansatz_theta}
\end{align}
Balancing units was not enough to determine the exponent of the
similarity solution for $\theta$, nor for finding the constants $a_0,
b_0$.
To find these we use ODE's (\ref{eq:ODEenergy}---\ref{eq:ODEint}).

Substituting (\ref{eq:ansatz_a}, \ref{eq:ansatz_b}) in equation (\ref{eq:ODEint}) gives the relation
\begin{equation}
  b_0 = \frac{a_0}{4}.\label{eq:a_b_rel}
\end{equation}
Further substitution of (\ref{eq:ansatz_a}--\ref{eq:a_b_rel}) into (\ref{eq:ODEenergy},\ref{eq:ODEmax}) yields
\begin{align*}
p(p+5)&=-\gamma,\\
\half p^2(2p+1)&=-\gamma,
\intertext{where $\gamma=\frac{\lambda}{4a_0}$.
Setting the two expressions for $-\gamma$ equal gives a polynomial equation for $p$:}
2p(p+5) &= p^2(2p+1).
\end{align*}

The solutions to this equation are $0, -2,\frac{5}{2}.$ 
Each solution for $p$ implies a solution for $\gamma$. 
By looking at equation (\ref{eq:evMax}) we see that $\gamma$ must be
non-negative. 
Indeed, when $s=1$, the LHS is non-negative, and $\mu=\lambda$ implies 
$\lambda\ge0$.
Since $a_0$ is positive,  $\gamma\ge0$.
The values $0, -2, \frac{5}{2}$ for $p$ imply the values $0, 6, -75$
for $\gamma$, respectively. 
Therefore, the only admissible values for $p$ are $0,-2$.
$p=0$ implies $\gamma = 0$, and this is physically non-interesting:
When $\gamma=0$, $\lambda=0$, hence the column has no weight. 
Either there is no gravity, or the density of the material is zero. 
We want to study the first positive eigenvalue $\lambda>0$. 
The solution $p=-2$ gives $\gamma = 6$ hence, $a_0 = \frac{\lambda}{24}$.
The similarity solution is found to be
\begin{align}
  \tilde a({s}) &= \frac{\lambda}{24} {s}^3,\label{eq:sim_a}\\
  \tilde b({s}) &= \frac{\lambda}{96} {s}^4,\\
  \tilde\theta({s}) &=  {s}^{-2}.\label{eq:sim_theta}
\end{align}
It is easy to see that the similarity solution
(\ref{eq:sim_a}--\ref{eq:sim_theta}) satisfies the BC
\eqref{eq:BC_at_tip} as $s\goto 0$.

\subsection{Peeling Away the Singularity}
While the similarity solution (\ref{eq:sim_a}--\ref{eq:sim_theta}) has
 the correct asymptotic behavior as ${s}\rightarrow 0$, it does not
satisfy BC \eqref{eq:BC_at_base} at the base ${s}=1$ (because
 $\tilde\theta(1)\ne 0$). 

We use the similarity solution to find a solution of the
 ODE that satisfies the BC at both ends.
Any solution can be written as a product of the similarity
solutions and new dependent variables $\alpha, \beta, \tau$:
\begin{align}
a &=  \tilde a \,\alpha\label{eq:def_alpha},\\
b &=  \tilde b \,\beta\label{eq:def_beta},\\
\theta &= \tilde\theta\, \tau\label{eq:def_tau}.
\end{align}
Substituting these expressions for $a, b$ and $\theta$ in the
differential equations (\ref{eq:ODEenergy}--\ref{eq:ODEint}) results in ODE's for $\alpha,\beta$ and $\tau$. 
These ODE's are homogeneous in ${s}$,
hence we use $t=-\ln {s}$ as the independent variable. 
The ODE for $\alpha(t), \beta(t), \tau(t)$ are the autonomous system (AS):
\begin{align}
(3-D)\brk{\alpha^2(2+D)\tau}-6\beta\tau&=0,\label{eq:AS_1}\\
(3+D)\brk{\alpha\Brk{(2+D)\tau}^2}-12\tau^2&=0,\label{eq:AS_2}\\
\brk{4-D}\beta-4\alpha&=0.\label{eq:AS_3}
\end{align}
In (\ref{eq:AS_1}--\ref{eq:AS_3}), $D$ denotes differentiation with respect to the variable $t$.
The BC (\ref{eq:BC_at_base},\ref{eq:BC_at_tip}) are also transformed.
At the base, $s=1,\,t=0$ the base BC \eqref{eq:BC_at_base} transform into
\begin{align}
 \beta(0) &= \frac{96}{\lambda},& \tau(0)&=0,\label{eq:BC_base}
\intertext{
At the tip, as $s\goto 0, t\goto\infty$, the BC \eqref{eq:BC_at_tip}
transform into}
\tau(t) \rightarrow &1 \text{ as } t\rightarrow\infty, & 
\beta(t)\rightarrow &1 \text{ as } t\rightarrow\infty.\label{eq:BC_tip}
\end{align}
The ODE's above, although written in an implicit form, can be rewritten as explicit expressions for $\tau_{tt}, \beta_t, \alpha_t$ as functions of $\tau, \tau_t, \beta, \alpha$. 
Note that due to the specific choice of dependent variables,
$\lambda$ is no longer a parameter in the AS.
It only appears in the BC at the base.
It is easy to see that $\alpha\equiv\beta\equiv 1, \tau \equiv 1$ is a critical point of the AS (\ref{eq:AS_1}--\ref{eq:AS_3}).
Setting $\alpha\equiv\beta\equiv\tau\equiv 1$ in
(\ref{eq:def_tau}--\ref{eq:def_beta}) recovers the similarity
solution (\ref{eq:sim_a}--\ref{eq:sim_theta}).

We seek a solution that satisfies the BC at $t=0$
and approaches the critical point as $t\rightarrow\infty$.
This solution satisfy all the BC.
A solution that approaches a critical point as its limit at $t=\infty$ is a
solution on the stable manifold of the critical point.

Since the critical point has both a stable an unstable manifold, it
is difficult to start on the stable manifold far from the
critical point and numerically follow the solution into the critical point.
The smallest numerical error will cause the solution to deviate on a
growing solution that does not approach the critical point.
Instead, we identify the stable manifold, and follow the solution
\emph{backwards} in $t$ from the critical point until the BC at $t=0$ are satisfied. 
This way, numerical errors will not be amplified. 
In fact, numerical errors will either decay or remain constant when
solving in this direction. 
The following section shows that the stable manifold is one
dimensional. 
Thus, there is only one direction in which to start the solution near
the critical point.
Solving the AS backwards from the critical point will follow the stable manifold. 
Since the BC at the base define a plane of co-dimension 1, the stable manifold is expected to intersect this plane.
Once the intersection point is found, the solution is fully determined.

\subsection{Stable Manifold}
To identify the stable and unstable manifolds of the
critical point, we linearize the AS around the critical point$(1, 1, 1)$, resulting in the linear ODE
\begin{equation}
\brk{\begin{matrix}
D(D+5)&0&D+3\\
D(D-1)&6&4(D-3)\\
0 &(D-4)&4
\end{matrix}}
\brk{\begin{matrix}\delta\tau\\ \delta\beta\\ \delta\alpha\end{matrix}}
=\brk{\begin{matrix}
0\\0\\0\end{matrix}}
\label{eq:linearized}
\end{equation}
Here, $(\delta\tau, \delta\beta, \delta\alpha)$ are the
deviations from $(1,1,1)$.
\eqref{eq:linearized} is a linear  system of ODE's, we therefore look for solutions of the form
\begin{equation*}
  (\delta\tau,  \delta\beta, \delta\alpha) =
  (\delta\tau_0, \delta\beta_0, \delta\alpha_0) e^{qt}.
\end{equation*}
Substituting this form of solution into equation \eqref{eq:linearized}
yields a generalized linear eigenvalue problem:
\begin{equation}
\brk{\begin{matrix}
q(q+5)&0&q+3\\
q(q-1)&6&4(q-3)\\
0 &(q-4)&4
\end{matrix}}
\brk{\begin{matrix}\delta\tau_0\\ \delta\beta_0\\ \delta\alpha_0\end{matrix}}
=\brk{\begin{matrix}
0\\0\\0\end{matrix}}.
\label{eq:linear_system}
\end{equation}

Equation \eqref{eq:linear_system} has non-trivial solutions for 4 different values of
$q: q_1=0, q_2=1, $\mbox{$q_3 \approx-5.5208$}, and \mbox{$q_4\approx6.5208$}. 
The solutions are spanned by the vectors given in the table below.
\begin{table}[h!]
 \begin{center}
 \begin{tabular}{>{$}l<{$}|>{$}r<{$} >{$}r<{$} >{$}r<{$} >{$}r<{$} }
   &S_1&S_2&S_3&S_4\\ \hline
\delta\tau_0&1&-2&0.876733&-0.126733\\
\delta\beta_0&0&3&0.420133&-1.5868\\
\delta\phi_0&0&4&1&1
\end{tabular}
\end{center}
\caption{possible solutions to equation \eqref{eq:linear_system}}
\label{tab:solutions}
\end{table}
As we are looking for the stable manifold, the only solution of
interest is the one with a negative $q$, i.e. solution $(q_3,S_3)$
%
Readers noticing the conspicuous form of solution $(q_2,S_2)$ are
referred to Appendix \ref{similarity:prop} for brief discussion of it.


To find the numerical solution of (\ref{eq:AS_1}--\ref{eq:AS_3}), we start near the stationary point, on the line tangent to the stable manifold, and solve the AS backwards in $t$.
The BC \eqref{eq:BC_base} at the base define a surface $B$ in the $\tau, \tau_t, \beta,
\alpha$ phase space:
\begin{align*}
B= \BRK{\brk{\tau, \tau_t, \beta, \alpha} \Big| \tau=0 \text{ and }
  \frac{96}{\lambda}-\beta=0}.
\end{align*}
The first condition, $\tau=0$, is used as a stopping condition the base
of the column.
The second, $\beta = \frac{96}{\lambda}$, identifies  $\lambda$ once
the base is found. 
Once $\lambda$ is found, the original functions, $\theta, a, b$ are
known from (\ref{eq:def_alpha}--\ref{eq:def_tau}).

\subsection{Numerical results}
\begin{figure}[htb]
 \center{\scalebox{0.75}{\includegraphics{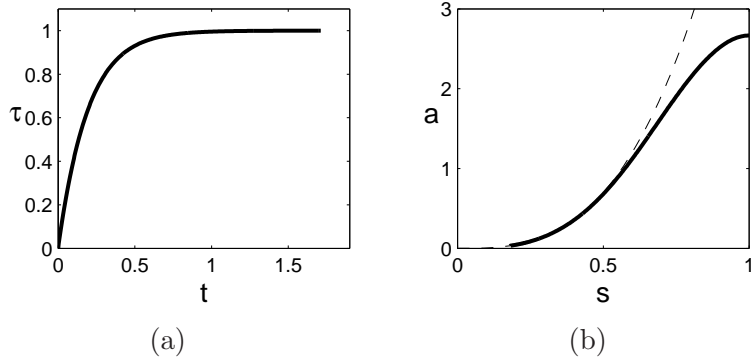}}\\
$\begin{array}{l@{\hspace{2in}}l}
\mbox{(a)}&\mbox{(b)}
\end{array}$
}
  \caption{The solutions of the tallest clamped column. Figure
    \ref{fig:a}(a) shows $\tau(t)$. 
    It  satisfies $\tau(0) = 0$ and as $t$ increases  it approaches $1$.
    Figure \ref{fig:a}(b) shows $a(s)$ (solid line) and the similarity
    solution (dashed line). 
    Since our solution for $\alpha(t)$ extends only up to a finite
    $t$, it isn't drawn for small values of $s$. 
    For small $s$ (i.e. large $t$) we can use the similarity solution
    to extend the solution to $s=0$.}
  \label{fig:a}
\end{figure}

Using \MATLAB's ODE solver \verb!ode45!, which uses the fourth-order Runge-Kutta method, the AS was solved. 
The solver was used with relative error tolerance of $10^{-4}$ and
an absolute error tolerance of $10^{-6}$.
The initial value for $(\tau, \tau_t, \alpha,\beta)$ was
\begin{equation}
(\tau, \tau_t, \alpha,\beta) = (1,0,1,1)+\delta \cdot(\tau_0,q\tau_0,\alpha_0,\beta_0).
\label{initial_cond}
\end{equation}
The values for $\tau_0, \alpha_0, \beta_0$ and $q$ are taken from $(q_3,S_3)$. 
Since the stable manifold is one dimensional there are two distinct
options for starting near the critical point. 
In terms of \eqref{initial_cond} the two options are choosing $\delta$ to be positive or negative.
The solution with positive $\delta$ fails to satisfy the BC at the
base. 
Geometrically, the trajectory is going in the wrong direction in the
phase space and does not intersect the plane $B$.
With $\delta=-0.0001$ the BC at the base are satisfied (that is, $\tau$ vanishes) at $\Delta t = -1.7114$. 
See figure \ref{fig:a}.

Since at the base, $\lambda = \frac{96}{\beta}$, the value for
$\lambda$ is given from the value of $\beta$. 
By this formula $\lambda = 134.1944$. 
Higher accuracy can be achieved by starting closer to the critical
point (smaller $\delta$), and setting the solver to a smaller tolerance.

Equations (\ref{eq:def_alpha}--\ref{eq:def_tau})
uniquely determine $\theta, a $ and $b$ given $\alpha, \beta, \tau$
and $\lambda$. 
Figure \ref{fig:a} shows the resulting cross-section $a(s)$ of the
$\lambda$--maximizing column tapering. 

\section{Other Boundary Conditions}
\label{sec:other_bc}
The fact that $\lambda$ appears only in the BC at the base has an
interesting implication for this problem: 
If we want to solve the
problem with a different BC at the base, we will have the same
analysis at the tip and the same similarity solution.
Since the stable manifold is one dimensional, we will have to start in
the same way while solving backwards.
The only difference will be when to stop. 

For instance, consider the problem of the tallest \emph{hinged}
column. A hinged column is a column that instead of being constrained to be
vertical at the base, is forced to have zero torque at the base. 

Note:
The lowest eigenvalue of the  hinged
column is zero.  
This is because at no gravity the column can be set at any angle.
To find the first non-trivial solution it would appear that we need to add
an additional orthogonality condition to make sure that we get the next
eigenfunction. 
A short calculation shows that the boundary conditions
and the ODE already guarantee the orthogonality of solutions to the trivial,
constant-angle, solution.

Formally, the difference is in the BC. In (\ref{eq:BC}), $\theta(1) = 0$
is replaced by the zero torque condition
\begin{align}
  a^2\,\theta_s &= 0 \text{ at } s=1.
  \label{eq:BC_2}
\intertext{The translation of this BC to  $\tau,\alpha, \beta$ is}
\alpha^2\,(2\tau+\tau_t) &= 0 \text{ at } t =0.\notag
\end{align}
The solution to this problem would be the design of the tallest
column that is hinged at the base and free at the top.
Using the same numerical method and the same initial conditions as before, a numerical solution was found.
After $\Delta t = -1.9470$, the BC, $\alpha^2(2\tau+\tau_t) = 0$, is
satisfied.

\begin{figure}[htb]
 \center{\scalebox{0.75}{\includegraphics{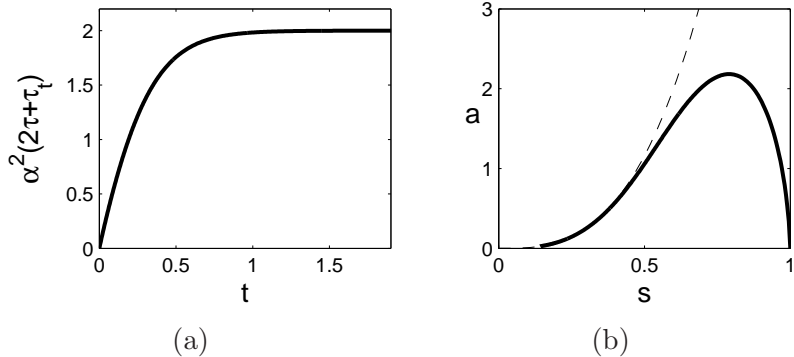}}}
$\begin{array}{l@{\hspace{2in}}l}
\mbox{(a)}&\mbox{(b)}
\end{array}$
  \caption{The solution of the tallest hinged column. 
    Figure  \ref{fig:a_2}(a) is the ``peeled'' torque,
    $\alpha^2(2\tau+\tau_t)$, as a function of $t$.
    It goes to zero at $t=0$ and since $\alpha,\,\tau$ go to the critical
    point as $t$ increases, it goes to 2.
    Figure \ref{fig:a_2}(b) shows $a(s)$ (solid line) and the similarity
    solution (dashed line). As before, $a(s)$ isn't drawn for small
    values of $s$. 
    For small $s$ (i.e. large $t$) we can use the similarity solution
    to extend the solution to $s=0$.}
  \label{fig:a_2}
\end{figure}

Again, since $\lambda = \frac{96}{\beta}$, we have the value of
$\lambda$. 
For these BC, the value is $\lambda = 222.7366$.
The cross-section and ``peeled'' torque are shown in Figure \ref{fig:a_2}.

The results predict a larger eigenvalue for the building with hinged base than with a clamped base.
This might be non-intuitive until one notices that the hinged column has within it a shorter clamped column. 
This column is also buckling and since it is shorter, its buckling load will be larger than the full length clamped beam.
This is also the case with simple loaded beams. 
The hinged loaded beam will be able to support a larger load than the clamped one. 
The reason for this non-intuitive result is the assumption that the column actually remains vertical. 

\section{Discussion}
\label{sec:dis}

The derivation of the equations above relies on the use of a variational derivative in order to find the cross-section $a(s)$ that produces a stationary eigenvalue $\lambda$. 
For this procedure to be valid, the operator, BVP (\ref{eq:BVP}, \ref{eq:BC}) should have discrete eigenvalues. 
As Cox and McCarthy have already shown, this is not always true. 
Nevertheless, the Keller and Niordson solution is very plausible. 
As shown in \cite{KellerTallest}, adding a small weight to the top of the column and letting the weight tend to zero produces a sequence of optimization problems.
These problems converge to the self-weight problem, and the solutions
(which in this case must exist) converge to the described solution.

We confined ourselves to demonstrating the use of the self similar solution in order to simplify the numerical solution of the given equations. 
Regardless of their derivation, the equations themselves require a solution and the method of peeling away the singularity using the similarity solution does away with the need of a iteration scheme.
This allows for any standard ODE solver to tackle the remaining problem. 
This method can be applied to other problems that have singularities. 
In a companion paper we employ a similar analysis to find the tapering of a javelin so that its lowest vibration mode has the highest possible frequency. 

\appendix
\section{Appendix}
\subsection{Specific Solutions in the Linearized System}
\label{similarity:prop}
We turn our attention to the conspicuous value of the eigenvalue
$q_2=1$ and respective eigenvector $S_2=(3,4,-2)$ which matches exactly
with the exponents of the similarity solution.
Here we show that this solution of the linearized ODE is to be
expected. 

Let 
\begin{align}
\dot{\Vec x}&=\Vec f(\Vec x)\label{eq:ODE}
\intertext{where}
\Vec x &= \brk{\begin{array}{c}x_1\\ x_2\\ \vdots\\ x_n\end{array}}\notag
\intertext{be an ODE. We further assume that }
  \Vec X(s) &= \brk{\begin{array}{c} a_1 s^{p_1}\\ a_2 s^{p_2} \\ \vdots \\a_n s^{p_n}\end{array}},\quad a_j\ne 0
 \label{eq:sim_sol}
\end{align}
is a solution to the ODE.
Any other solution $\Vec Y$ can be written as
\begin{equation}
\Vec Y(s) = A(s)\, \Vec X(s)\label{eq:Y}
\end{equation}
where $A(s)$ is a diagonal matrix.
The ODE that $A(s)$ must satisfy can be found by substituting
(\ref{eq:Y},\ref{eq:sim_sol}) into \eqref{eq:ODE}.
\begin{equation*}
\Vec f(A(s)\Vec X(s)) = \Vec f(\Vec Y(s))=\dot{\Vec Y}(s) = \dot{A}(s)\Vec X(s) + A(s)\dot{\Vec X}(s) =
\dot{A}(s)\Vec X(s)+A(s)P\Vec X(s)s^{-1}.
\end{equation*}
Here, $P$ is a diagonal matrix with $P_{ii}=p_i$.
In other words, $A(s)$ must satisfy
\begin{equation}
\Vec f(A(s)\Vec X(s))= \dot{A}(s)\Vec X(s)+A(s)P\Vec X(s)s^{-1}.\label{eq:A_ODE}
\end{equation}
Note that $A(s) \equiv I$ is a solution of this system.
Therefore, $I$ is a critical point of the system for $A$.
The stable and unstable manifolds of this system are important for the
type of solution method described in this paper. 
The following proposition shows the existence of a specific unstable
direction about the critical point $I$.

\begin{prop*}
The linearization of \eqref{eq:A_ODE} around $I$ has a solution
\end{prop*}
\begin{equation}
  \delta A(s) = Ps^{-1}.\label{eq:prop1}
\end{equation}
In the scaled variable, $t=-\ln s$, this corresponds to an eigenvalue
of 1 and an eigenvector $P$.

\begin{proof}
The linearization of \eqref{eq:A_ODE} is 
\begin{align}  
  J\!\Vec f\, \delta A(s)\Vec X(s)&=\dot{\delta A}(s)\Vec X(s)+\delta A(s)P\Vec X(s)s^{-1}.\label{eq:Lin_A_ODE}
  \intertext{Here, $\delta A$ is a diagonal matrix (the perturbation from the solution $I$) and  $J\!\Vec f$ is the Jacobian matrix of $\Vec f$ evaluated at $A(s)\Vec X(s)$.
    To find out more about $J\!\Vec f$, we differentiate the equality}
  Ps^{-1}\Vec X &= \Vec f(\Vec X) \label{eq:X_sol}
  \intertext{with respect to $s$:}
  P(P-I)\Vec X(s)s^{-2} &= J\!\Vec f\, P\Vec X(s)s^{-1}.\label{eq:diff}
  \intertext{Equation \eqref{eq:X_sol} is derived by substituting \eqref{eq:sim_sol} into \eqref{eq:ODE}.
To show that $\delta A(s) = Ps^{-1}$ solves equation \eqref{eq:Lin_A_ODE}, we
    substitute \eqref{eq:prop1} into \eqref{eq:Lin_A_ODE}:}
  -Ps^{-2}\Vec X(s)+P^2\Vec X(s)s^{-2}&=J\!\Vec f\,P\Vec X(s)s^{-1}.\notag\\
  \intertext{From equation \eqref{eq:diff} we substitute for the RHS}
  -Ps^{-2}\Vec X(s)+P^2\Vec X(s)s^{-2}&=P(P-I)\Vec X(s)s^{-2}\notag
\end{align}
This shows that indeed $\delta A(s) = P s^{-1}$ is a solution of the
linearized ODE. 
\end{proof}

\subsection{Calculation of the Lagrange multiplier}
\label{calc:mu}

To calculate the value of $\mu$ in (\ref{eq:evMax}), we multiply the
equation by $a(s)$ and integrate from $0$ to $1$:
\begin{align}
\int_0^12a^2\,\theta_s^2 \,ds -\lambda \int_0^1a\int_s^1\theta^2(t)\,dt \,ds &=\mu\int_0^1\theta^2(r)\int_0^ra(t)\,dt\,dr \int_0^1a\,ds . \notag
\intertext{
Since the RHS of \eqref{eq:evMax} is a constant and $\int_0^1a(s)\, ds =1$, the RHS stays unchanged.
In the LHS, integrating the first term by parts, changing the order of
integration of the second term and using the BC gives}
-2\int_0^1\brk{a^2\,\theta_s}_s\theta \,ds -\lambda \int_0^1\theta^2\int_0^sa(t)\,dt \,ds &=\mu\int_0^1\theta^2(r)\int_0^ra(t)\,dt\,dr.\notag
\intertext{Lastly, using (\ref{eq:linBVP}) we can see that}
2\lambda\int_0^1\theta^2\int_0^sa(t) \,dt\,ds -\lambda \int_0^1\theta^2\int_0^sa(t)\,dt \,ds &=\mu\int_0^1\theta^2(r)\int_0^ra(t)\,dt\,dr\notag
\end{align}
Therefore,  $\lambda =\mu$.

\subsection{Torque Balance }
\label{app:torque}

\begin{figure}[th]
\begin{center}

\scalebox{0.6}{\includegraphics{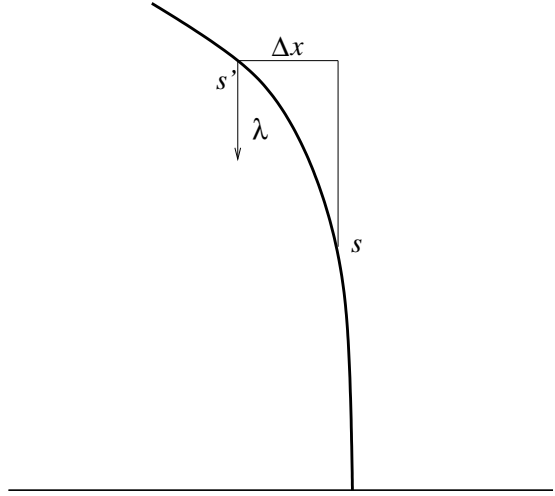}}

\caption{The torque at point $s$ is due to the sum of the torques from points $s'<s$ along the column. 
Each point creates a torque $a(s')\,\lambda\,\Delta x \,ds'$ (recall that $\lambda$ is the non-dimensional gravity). 
The curvature at the point $s$ is proportional to this torque.}
\label{fig:torque}
\end{center}
\end{figure}

The differential equation \eqref{eq:BVP} is a local torque balance.
Integrating equation \eqref{eq:BVP} and using the second of BC
\eqref{eq:BC} gives 
\begin{equation}
   a^2(s)\,\theta_s(s) = \lambda\int_0^s \sin \theta\fbrk{s'}\int_{0}^{s'}a\fbrk{t}\,dt\,ds'.\label{eq:int_BVP}
\end{equation}
Changing the order of integration in the RHS of \eqref{eq:int_BVP} we see that
\begin{equation}
   a^2(s)\,\theta_s(s) = \lambda\int_0^s a\fbrk{s'}\int_{s'}^{s}\sin \theta\fbrk{t}\,dt\,ds'.\label{eq:int_BVP_switch}
\end{equation}
The inner integral is the horizontal displacement $\Delta x$ of the point $s'$
from the point $s$.
The sections of the column above $s$ exert a gravity generated torque
on the point $s$.
This is the RHS of \eqref{eq:int_BVP_switch}.
The LHS expresses the elastic response to this torque: The curvature
$\theta_s$ of the beam at $s$ is proportional to the imposed torque.
See figure (\ref{fig:torque}).

\bibliographystyle{amsplain}
\bibliography{general}

\end{document}